\documentclass{article}

\usepackage[english]{babel}
\usepackage[utf8]{inputenc}
\usepackage[T1]{fontenc}
\usepackage{natbib}
\setcitestyle{authoryear, open={(},close={)}}

\usepackage{fullpage}

\usepackage{graphicx}
\usepackage[colorinlistoftodos]{todonotes}
\usepackage{multirow}
\usepackage{booktabs}
\usepackage[colorlinks=true, allcolors=blue]{hyperref}
\usepackage[shortlabels]{enumitem}
\usepackage{amsthm,amsmath,bm,bbm}
\usepackage{amssymb,mathtools}
\allowdisplaybreaks[1]

\usepackage{listings}
\usepackage{parskip}
\newlength\tindent
\usepackage{caption}
\usepackage{subcaption}

\usepackage{appendix}

\def\independent{\perp\!\!\!\perp}

\def\E{\textup{E}}
\def\P{\textup{P}}

\def\logit{\text{logit}}
\def\expit{\text{expit}}

\title{Multiple imputation for propensity score analysis with covariates missing at random: some clarity on \textit{within} and \textit{across} methods}

\author{Trang Quynh Nguyen\footnote{Please address correspondance to the first author at trang.nguyen@jhu.edu.}~~and Elizabeth A. Stuart\\~\\Johns Hopkins Bloomberg School of Public Health, Baltimore, MD, USA}

\begin{document}

\maketitle

\begin{abstract}
    \noindent In epidemiology and social sciences, propensity score methods are popular for estimating treatment effects using observational data, and multiple imputation is popular for handling covariate missingness. However, how to appropriately use multiple imputation for propensity score analysis is not completely clear. This paper aims to bring clarity on the consistency (or lack thereof) of methods that have been proposed, focusing on the \textit{within} approach (where the effect is estimated separately in each imputed dataset and then the multiple estimates are combined) and the \textit{across} approach (where typically propensity scores are averaged across imputed datasets before being used for effect estimation). We show that the \textit{within} method is valid and can be used with any causal effect estimator that is consistent in the full-data setting. Existing \textit{across} methods are inconsistent, but a different \textit{across} method that averages the inverse probability weights across imputed datasets is consistent for propensity score weighting. We also comment on methods that rely on imputing a function of the missing covariate rather than the covariate itself, including imputation of the propensity score and of the probability weight. Based on consistency results and practical flexibility, we recommend generally using the standard \textit{within} method. Throughout, we provide intuition to make the results meaningful to the broad audience of applied researchers.
    
    \bigskip
    
\end{abstract}

\section{Introduction}

Consider the common setting of an observational study where the analyst is interested in conducting a propensity score analysis. 
Here
$A$ is a binary exposure, $Y$ is the observed outcome, and $Y_1$ and $Y_0$ are the potential outcomes \citep{rubin1974EstimatingCausalEffects} under exposure (treatment) and non-exposure (control condition). 
We take the estimand to be the average treatment effect (ATE), defined as $\E[Y_1-Y_0]=\E[Y_1]-\E[Y_0]$ (but this discussion is also relevant to other average effects such as the average treatment effect on the treated or the average treatment effect on another target population).
We make the usual causal inference assumptions: (i) \textit{SUTVA} (i.e., no interference and treatment variation irrelevance) \citep{rubin1980RandomizationAnalysisExperimental}, which implies $Y=AY_1+(1-A)Y_0$; (ii) \textit{unconfoundedness} \citep{imbens2008RubinCausalModel} given covariates, formally, $A\independent Y_a\mid Z,X$ for $a=0,1$ (where $Z$ and $X$ denote covariates, to be differentiated shortly); and (iii) \textit{positivity} \citep{rosenbaum1983CentralRolePropensity}, formally, $\epsilon<\P(A=1\mid Z,X)<1-\epsilon$ for some $\epsilon>0$.

The challenge is that while $Z,A,Y$ are fully observed, there are missing values in $X$, which complicates the analysis. To make progress the analyst will need to make some assumption about the missingness mechanism.
Here we consider the common \textit{missing at random} (MAR) assumption: the missingness depends on the observed data, not on the missing values. 
While $Z$ and $X$ are typically multivariate, for the sake of exposition we will focus on the univariate $X$ case (but all the issues discussed here and all our conclusions apply generally to multivariate $X$ unless otherwise noted). With univariate $X$, MAR simply means $R\independent X\mid Z,A,Y$, where $R$ is the binary indicator for observing $X$. That is, whether $X$ is missing or not may depend on the other covariates and/or on the treatment condition and/or the outcome, but not on the value of $X$ itself. In this setting complete-case analysis generally results in biased ATE estimation (due to distortion of the joint covariate and conditional outcome distributions), so the missingness needs to be handled with care.

Popular approaches to handling missing data assuming MAR include multiple imputation (MI) \citep{rubin1987multiple} and inverse probability of response weighting \citep{horvitz1952GeneralizationSamplingReplacement}. In this paper we focus on MI, which is flexible for handling missing covariates.
We aim to bring some clarity on the \textit{consistency} (i.e., the property of yielding estimates that are close to the truth in large samples%
\footnote{The technical definition is that as sample size approaches infinity, the estimator converges in probability to the true value.}) or lack thereof of several methods to combine MI with propensity score analysis that have been explored in the literature.
Among these, two extensively considered by methodologists \citep{hill2004ReducingBiasTreatment,mitra2016ComparisonTwoMethods,leyrat2019PropensityScoreAnalysis,ling2021HowApplyMultiple,leite2021ImputationMissingCovariate} are the \textit{within} and \textit{across} methods \citep[terms coined by][]{mitra2016ComparisonTwoMethods}. Both involve imputing the missing covariates; the \textit{within} method implements the full propensity score analysis in each imputed dataset and pools treatment effect estimates, whereas the \textit{across} method averages the estimated propensity scores across imputed datasets and uses the set of average propensity scores to obtain a single treatment effect estimate. A variant of the \textit{across} method \citep[see][]{leyrat2019PropensityScoreAnalysis,ling2021HowApplyMultiple} averages the parameters of the propensity score model as well as the imputed covariate values across imputed datasets, and uses these to compute a single set of propensity scores. And a completely different strategy is to impute the propensity score, explored in \cite{ling2021HowApplyMultiple}.

Before discussing these various methods, a general note about the MI part: we presume that the imputation model includes all analysis variables so as to avoid uncongeniality with the analysis model \citep{meng1994MultipleimputationInferencesUncongenial}. Our discussion thus excludes some early investigations that do not use outcome data in imputing covariates in order to preserve the separation of the \textit{design} and \textit{analysis} steps in an observational study \citep{rubin2007DesignAnalysisObservational};  these include \cite{mitra2016ComparisonTwoMethods}  and part of \cite{hill2004ReducingBiasTreatment}. Model uncongeniality generally results in bias because the imputation distorts the joint distribution of the analysis variables; in the current setting it turns out that not including the outcome in the imputation model is equivalent to assuming the covariate to be imputed is not a confounder \citep{seaman2014InverseProbabilityWeighting}. 
As for \textit{design}-\textit{analysis} separation concerns, we propose that one treat MI as an early step (step 0) that can be kept separate from both \textit{design} (step 1, which typically involves covariate balancing) and \textit{analysis} (step 2, which estimates effects using balanced data) -- rather than thinking of it as part of the design step.
One way to ensure MI-design-analysis separation, for example, is to impute separately in each treatment group and split data access accordingly; then the data used for MI ($Z,X$ and $Y$, but one group at a time) allows neither covariate balancing nor effect estimation.

We will first review existing results about methods for combining MI with propensity score analysis.

\subsection{What is known about these methods}

The \textit{within} method and the original \textit{across} method (that averages propensity scores) have been investigated and compared most often. (Unless a specific benchmark is referenced, all bias reduction mentioned here is compared to complete-case analysis.) For the most part simulation studies suggest that both methods reduce bias. Which of the two should be preferred, and whether that differs between propensity score matching and weighting is less clear.
For propensity score matching, \cite{hill2004ReducingBiasTreatment} find that the \textit{within} method performs much better than the \textit{across} method.
\cite{devries2016CommentsPropensityScore} note that 
method performance with matching may be affected by positivity violation;
when handling positivity is carefully, they find that both methods reduce bias, with the \textit{within} method performing slightly better.
For propensity score weighting, \citeauthor{leyrat2019PropensityScoreAnalysis}'s (\citeyear{leyrat2019PropensityScoreAnalysis}) simulations find that the \textit{within} method is unbiased while the \textit{across} method is biased (although both reduce bias substantially). In \cite{ling2021HowApplyMultiple}, on the other hand, which method performs better varies across data generating mechanisms.
\cite{leite2021ImputationMissingCovariate} find that the \textit{across} method reduces bias substantially
(but do not compare the two methods). In \cite{granger2019AvoidingPitfallsWhen} the \textit{within} method outperforms the \textit{across} method for both propensity score matching and weighting.

Regarding the other methods, two studies (both using propensity score weighting) report different results on the \textit{across} method variant that averages propensity score model parameters and imputed $X$ values: in \cite{leyrat2019PropensityScoreAnalysis} it reduces bias slightly (relative to the original \textit{across} method), while in \cite{ling2021HowApplyMultiple} it increases absolute bias. The latter study also reports that propensity score imputation strategies perform poorly, but this investigation suffers from several flaws (see section \ref{sec:imputePS}) that cast doubt on this result.

In brief, simulation results seem to favor the \textit{within} method, but there are some mixed or inconclusive results. However, simulation results may be hard to assess as they may be affected by many issues specific to the propensity score method or the MI. Positivity violation, noted above, for example, affects both propensity score matching and weighting -- resulting in poor matches or large weights. Imputation models may be hard to specify correctly, 
resulting in biased imputed values. It is thus hard to judge whether a worse performance is due to one of these challenges or reflects an intrinsic property of the method. We should therefore also rely on theoretical reasoning.

Two theoretical results exist for the \textit{within} method combined with propensity score \textit{weighting}. \cite{seaman2014InverseProbabilityWeighting} prove that this method is consistent -- if the imputation and propensity score models are correctly specified and the number of imputations $M$ is infinite. The proof manipulates the estimating equation corresponding to the full-data estimator using iterated expectation and swaps out the distribution of $X$ given $(Z,A,Y)$ for the asymptotically equivalent imputation distribution, showing that the imputation-based estimator converges to the same estimand. This proof may be inaccessible to many applied users, as it is mathematically succinct and is not accompanied with an intuitive explanation. 
Later \cite{leyrat2019PropensityScoreAnalysis}, after citing \cite{seaman2014InverseProbabilityWeighting}, offer another proof with an intuitive explanation. Key to this proof are two claims that (under the same conditions as in \cite{seaman2014InverseProbabilityWeighting}) the propensity score model is the same in the imputed datasets as in the original data; and unconfoundedness holds in the imputed datasets. These combined imply that the expectations of group-specific inverse probability weighted outcome means in the imputed datasets are equal to $\E[Y_1]$ and $\E[Y_0]$.
Unfortunately, the second claim is not correct (see section \ref{sec:within}). This leaves us still without a simple and intuitive explanation for applied users.

As for the \textit{across} method and its variant mentioned above, \cite{leyrat2019PropensityScoreAnalysis} point out that since the propensity score estimates obtained from these methods are not \textit{finer} \citep[in the sense of][]{rosenbaum1983CentralRolePropensity} than the full-data propensity score $e(X,Z):=\P(A=1\mid Z,X)$, it is not a balancing score and thus does not result in consistent $\tau$ estimation. This is the best explanation we found in the literature on these methods. To grasp this, however, requires a good understanding of the technical content of the seminal \cite{rosenbaum1983CentralRolePropensity} paper.

\subsection{Our contribution}

In this paper we provide a simple explanation for the consistency of the \textit{within} method (under the same conditions considered by previous authors). In addition, we argue that this result applies more generally, not restricted to the inverse probability weighting estimator; this is broader than what is known from \cite{seaman2014InverseProbabilityWeighting}. With the \textit{across} methods, we complement the high level explanation of \cite{leyrat2019PropensityScoreAnalysis} with a tangible understanding of the inconsistency problem
by zooming in to what is specifically estimated by averaging propensity scores (or averaging model parameters and imputed values) across imputed datasets and viewing the situation through a measurement error lens.
We note a special case where for propensity score weighting, a different \textit{across} method that averages the inverse probability weights (not the propensity scores) is consistent; this method is closely related to the \textit{within} method. 
Separately, considering the approach of imputing a function of the missing covariate rather than the covariate itself, we correct some issues with the propensity score imputation method considered in \cite{ling2021HowApplyMultiple} and point out that a \textit{within} method using \textit{correct} propensity score imputation is consistent. In addition, we note that mean (instead of stochastic) imputation of the probability weight works for propensity score weighting analysis.

With the current focus on consistency, the discussion here applies equally to imputation using a Bayesian model (standard in popular MI software) and imputation based on a model fit with maximum likelihood.

This paper does not cover variance of MI-based estimators or the estimation of such variance, which deserve dedicated investigation. We will briefly comment on this at the end of the paper.

\section{Assumptions}

We adopt the same assumptions used in \cite{seaman2014InverseProbabilityWeighting} and \cite{leyrat2019PropensityScoreAnalysis}, that the imputation model and the propensity score model are correctly specified and regularity conditions hold for them to converge to the truth. With \textit{across} methods, we consider the behavior of the methods as $M\to\infty$, where Monte Carlo integration becomes exact and we do not lose any information from the imputation model. These elements combined allow us to focus our attention to the ideal (asymptotic) setting where (i)~imputed values are generated from the same distribution as the true distribution of $X$ given $(Z,A,Y)$, and (ii)~averaging over the imputed datasets means integrating out the missing values by averaging over this distribution. This avoids complicating presentation with extraneous details, and is enough for the sake of arguing consistency or lack thereof.

\section{The \textit{within} method}\label{sec:within}

In the \textit{within} method, missing data are imputed, effect estimation by propensity score analysis is then conducted separately on each imputed dataset, and finally effect estimates are averaged over imputed datasets.

To differentiate an imputed value from the true $X$, we use the notation $X^*$. Let $X^\dagger=RX+(1-R)X^*$, so $X^\dagger$ is the covariate in place of $X$ in an imputed dataset. 

The reason the \textit{within} method works is quite simple. The first part of the argument is standard causal inference reasoning, putting aside the issue of missing data. Here the usual causal inference assumptions (SUTVA, unconfoundedness and positivity) equate the ATE (a contrast of potential outcome means) to $\tau:=\E\{\E[Y\mid Z,X,A=1]-\E[Y\mid Z,X,A=0]\}$, a function of the joint distribution of $(Z,X,A,Y)$ (which does not involve potential outcomes).
Without missing data, the ATE could be estimated by estimating $\tau$.

The challenge then is to estimate this parameter $\tau$ while $X$ is not fully observed. This is where imputation comes into play. With correct specification, the imputation model (at the limit as $n\to\infty$) captures the true distribution of $X$ given $(Z,A,Y)$. This implies that $X^*$ (and thus $X^\dagger$) is distributed the same as $X$ given $(Z,A,Y)$, which implies that the joint distribution of $(Z,X^\dagger,A,Y)$ in the imputed dataset is the same as the joint distribution of $(Z,X,A,Y)$ in the real data, formally, $\P(Z,X^\dagger=x,A,Y)=\P(Z,X=x,A,Y)$. Intuitively, imputation using the correct model recovers the true joint distribution of the analysis variables. It follows that $\E\{\E[Y\mid Z,X^\dagger,A=1]-\E[Y\mid Z,X^\dagger,A=0]\}=\tau$.

A metaphor for this second part of the argument is a special kind of multiverse. The original dataset is the universe we are in,
where we do not see everything, as some parts (the missing $X$ values) are masked from us. MI takes us to alternate 
universes where we see everything, where everything we saw in our original universe stays exactly the same, but everything not originally observed vary. (The truth is, only one universe (the one we are in) is real; the other ones are all imagined and just different guesses of what the unobserved part of the real universe might be.) The key point is that all these universes, real and imagined, obey the same law (the joint $(Z,X,A,Y)$ distribution). $\tau$ is simply a feature of this law, thus is shared across the universes.

That the same $\tau$ is shared by the $(Z,X^\dagger,A,Y)$ distribution and the $(Z,X,A,Y)$ distribution justifies using each imputed dataset to estimate $\tau$.
What this means is that the \textit{within} method is consistent for any base estimator that would have been consistent had there been no missing data. This is not restricted to propensity score weighting \citep[the case that has been proven by][]{seaman2014InverseProbabilityWeighting}, but also applies to matching and other covariate balancing as well as outcome regression and doubly robust methods -- provided those methods are consistent for the case with full data.

Our argument for the consistency of the \textit{within} method here differs from the argument in \cite{leyrat2019PropensityScoreAnalysis}, which relies on the proposition that unconfoundedness holds in the imputed dataset (i.e., $A\independent Y_a\mid Z,X^\dagger$ for $a=0,1$) to justify using the imputed dataset to estimate the treatment effect. This claim is incorrect, however, due to an error in the proof (see explanation in Appendix \ref{app:A}). Intuitively, this claim essentially implies that the imputation recovers the joint distributions of $(Z,X,A,Y_a)$ for $a=0,1$. This cannot be the case, however, as the imputation of $X$ is conditional on $(Z,A,Y)$; it is informed by $Y$, not $Y_1$ or $Y_0$. Our argument, on the other hand, only relies on the imputation recovering the distribution of $(Z,X,A,Y)$, as the parameter we need to estimate, $\tau$, is agnostic of potential outcomes.

Side note: As the key point is to recover the $(Z,X,A,Y)$ distribution, it is important to include all the variables $Z,A,Y$ as predictors in the imputation model, so that $X^*$ is generated based on $(Z,A,Y)$. This reiterates the need to use the outcome in imputing covariates.

Even though their argument does not hold, \citeauthor{leyrat2019PropensityScoreAnalysis}'s point about recovering unconfoundedness is interesting. There is indeed a different explanation for why the \textit{within} method works that involves imagining (but not implementing) a more complex imputation scheme that recovers the joint distributions of $(Z,X,A,Y_a)$ for $a=0,1$, and thus recovers unconfoundedness (see Appendix \ref{app:B}).

\section{The \textit{across} methods}

We will revisit the two existing \textit{across} methods (aPS and aPM) which are not consistent, then consider a different \textit{across} method (aPW) that is consistent for propensity score weighting.

These \textit{across} methods do not involve searching for $\tau$ in each of the multiple universes. Rather, we take a quick flight across the multiverse to learn something and bring it back and hope to use it to find $\tau$ in our universe. Roughly speaking, this is how we define the class of \textit{across} methods.

\subsection{Averaging the propensity score (aPS)}

In the original \textit{across} method, the something we learn flying across the multiverse is the average of the propensity scores for each individual, which we refer to as the \textit{average propensity score} (aPS).
To understand this method, we ask what is being estimated by the aPS.

Since the imputation recovers the $(Z,X,A,Y)$ joint distribution, it recovers the distribution of exposure given covariates, formally, $\P(A\mid Z,X=x)=\P(A\mid Z,X^\dagger=x)$ for $x$ values in the support of $X$ given $Z$. This means the propensity score model fit to the imputed dataset estimates the true propensity score function $e(z,x)$. 

This implies that for each unit with observed $X$, the propensity score in each imputed dataset is consistent for the true $e(Z,X)$, so the aPS is also consistent for the true $e(Z,X)$.
What is perhaps less obvious is what is estimated by the aPS for a unit with missing $X$. Because the imputed $X^*$ has been generated conditional on $(Z,A,Y)$ using a distribution that is the same as the distribution of $X$ given $(Z,A,Y)$, averaging across imputed datasets means averaging over this distribution. The aPS is thus an estimate of $\E[e(Z,X)\mid Z,A,Y]$.
Had the ultimate goal been to estimate the propensity score for each unit, this would be the best we can do: for units with observed $X$ we have consistent estimation of the propensity score, and for units missing $X$ we have consistent estimation of the projection of the propensity score on observed data $(Z,A,Y)$. However, estimating propensity scores is just an intermediate step for estimation of $\tau$.

For $\tau$ estimation (via matching or weighting or any other method that uses propensity scores), we now have a situation of differential measurement error: we have estimates of $e(Z,X)$ for units with observed $X$, of $\E[e(Z,X)\mid Z,Y,A=1]$ for units missing $X$ in the treated group, and of $\E[e(Z,X)\mid Z,Y,A=0]$ for units missing $X$ in the control group. For an easy analogy,  take matching for example (but this measurement error affects all methods). Let us think of $e(Z,X)$ values as apples of varying sizes, and the ideal is to match apples with apples of the same size. However, mixed in with the apples there are peaches ($\E[e(Z,X)\mid Z,Y,A=1]$ values) and oranges ($\E[e(Z,X)\mid Z,Y,A=0]$ values). This means while we are still matching by size, sometimes instead of matching apples to apples, we end up matching apples to oranges, or apples to peaches, or oranges to peaches. As a result, the aPS \textit{across} method is not consistent for $\tau$ estimation.%
\footnote{That this measurement error problem affects propensity score weighting can be recognized by seeing weighting as a form of group matching, i.e., creating two weighted groups with the same propensity score distribution (the same distribution of apples in our analogy). Measurement error (mixing of peaches and oranges with the apples) results in groups that are mis-matched.}

\subsection{Averaging model parameters (aPM)}

This specific method applies to parametric propensity score models, such as the logit model. Instead of averaging propensity scores, here one averages the estimated parameters of the model over the imputed datasets to form a pooled propensity score model, which we refer to as the \textit{average parameters model} (aPM). To compute propensity scores using this model, the method investigated in \cite{leyrat2019PropensityScoreAnalysis} and \cite{ling2021HowApplyMultiple} replaces missing $X$ with the average of the imputed values $X^*$ for the unit (denoted $\bar X^*$, which estimates $\E[X\mid Z,A,Y]$). We discuss these two steps separately.

The fact that all propensity score models in imputed datasets estimate the same $e(z,x)$ function justifies pooling these fitted models in some way to obtain a combined estimate of the $e(z,x)$ function. Note that with the aPS method above, the averaging of propensity scores for units with observed $X$ in fact is a simple way to obtain a combined estimate of the $e(z,x)$ function -- evaluated at the observed $(Z,X)$ values. 
Now consider the parameter averaging strategy. For the logit model, averaging estimated coefficients across imputed datasets amounts to averaging the estimated logits, which is consistent for the true logit function, so the aPM is consistent for the true propensity score function $e(z,x)$. 
For units with observed $X$, this model provides consistent propensity score estimates.

As for the use of $\bar X^*$ in computing the propensity score for a unit with missing $X$, that amounts to replacing that unit's propensity score with the propensity score of a unit whose $Z$ value is the same but whose $X$ value is equal to the average $X$ of units with the same $(Z,A,Y)$ values.
If the model is logit-linear in $X$, this turns out to be equivalent to averaging the logit propensity score over the distribution of $X$ given $(Z,A,Y)$, which means the $\bar X^*$-based propensity score is an estimate of $\expit\big(\E\{\logit[e(Z,X)]\mid Z,A,Y\}\big)$. 
So similar to the aPS method, here we also have a differential measurement error situation: we have estimates of $e(Z,X)$ for units with observed $X$, of $\expit\big(\E\{\logit[e(Z,X)]\mid Z,Y,A=1\}\big)$ for treated units missing $X$, and $\expit\big(\E\{\logit[e(Z,X)]\mid Z,Y,A=0\}\big)$ for control units missing $X$.
As a result, the aPM \textit{across} method is not consistent.

To sum up, both the aPS and aPM methods (and more generally methods that aim to create \textit{one set of estimated propensity scores}) are not consistent because they result in a case of measurement error on the propensity score, where the error structure varies depending on whether $X$ is missing or observed, and if $X$ is missing, whether the unit is in the treated or control group. This speaks directly to \citeauthor{leyrat2019PropensityScoreAnalysis}'s (\citeyear{leyrat2019PropensityScoreAnalysis}) point: neither method provides a balancing score. Here we see clearly how that is so.

\subsection{Averaging the probability weight (aPW)}

If the base method is propensity score weighting, a different \textit{across} method actually works. With this method, one would compute the inverse probability weights in the imputed datasets and average them for each unit over the imputed datasets, and then use these \textit{average probability weights} (aPW) in the estimation of $\tau$.

Let $\omega(Z,X,A)=\frac{A}{e(Z,X)}+\frac{1-A}{1-e(Z,X)}$ denote the inverse probability weight. The aPW estimates different quantities depending on whether $X$ is observed: for units with observed $X$ it estimates the true probability weight $\omega(Z,X,A)$; for units missing $X$ it estimates $\E[\omega(Z,X,A)\mid Z,A,Y]$, which we denote by $\tilde\omega(Z,A,Y)$. 
Note that among the \textit{across} methods, aPW produces the best guess of the true weight $\omega(Z,X,A)$.



This best guess turns out to be a good proxy for the true weight in the sense that it does not bias $\tau$ estimation.
Its justification is based on the full-data identification formula that justifies propensity score weighting: 
$$\tau=\E\left[\frac{A}{e(Z,X)}Y-\frac{1-A}{1-e(Z,X)}Y\right]=\underbrace{\textcolor{blue}{\E\left[A\,\omega(Z,X,A)Y\right]}}_{\textstyle=:\theta_1}\,-\,\underbrace{\E\left[(1-A)\omega(Z,X,A)Y\right]}_{\textstyle=:\theta_0}.$$
If suffices to consider one of the two terms, $\theta_1$. Using iterated expectation conditioning on $(Z,A,Y)$ obtains
$$\theta_1=\E\big\{\E\left[A\,\omega(Z,X,A)Y\mid Z,A,Y\right]\big\}=\E\big\{A\,\E\left[\omega(Z,X,A)\mid Z,A,Y\right]Y\big\}=\textcolor{blue}{\E[A\,\tilde\omega(Z,A,Y)Y]}.$$
This justifies replacing the inverse probability weight $\omega(Z,X,A)$ in the $\tau$ formula with $\tilde\omega(Z,A,Y)$.
This simple proof unfortunately does not seem to offer, at least to us, any immediate intuition.


A perhaps easier way to see that propensity score weighting using the aPW \textit{across} method is consistent is to note that it is equivalent, or asymptotically equivalent, to the \textit{within} method (which we know is consistent). Again, we will just consider the estimation of the $\theta_1$ term. There are two forms of estimation by inverse probability weighting: the \cite{hajek1971comment} estimator is a weighted average of the outcome, while the \cite{horvitz1952GeneralizationSamplingReplacement} (H-T) estimator multiplies the outcome of each unit by the weight and sums over all units but then divides the sum by the sample size. These estimators of $\theta_1$ using imputed dataset $k$ are
$$\hat\theta_{1,\text{H\'ajek-}k}:=\frac{\sum_{i=1}^Nw_{1ik}Y_i}{\sum_{j=1}^Nw_{1jk}}~~\text{and}~~\hat\theta_{1,\text{HT-}k}:=\frac{1}{N}\sum_{i=1}^Nw_{1ik}Y_i,$$
where $w_{1ik}:=\frac{A_i}{\hat e(Z_i,X_{ik}^\dagger)}$ is the estimated weight for estimating $\theta_1$ of unit $i$ in imputed dataset $k$.
Denote the average of these weights of unit $i$ over the imputed datasets by $\bar w_{1i}:=\frac{1}{M}\sum_{k=1}^Mw_{1ik}$.

When using the H-T as the base estimator, the aPW \textit{across} method and the \textit{within} method give the two estimators
$$\hat\theta_{1,\text{HT-aPW}}:=\frac{1}{N}\sum_{i=1}^N\bar w_{1i}Y_i~~\text{and}~~\hat\theta_{1,\text{HT-within}}:=\frac{1}{M}\sum_{k=1}^M\hat\theta_{1,\text{HT-}k},$$
which are exactly equivalent as they are equal to the same linear combination of the units' outcomes,
$$\sum_{i=1}^N\frac{\sum_{k=1}^Mw_{1ik}}{NM}Y_i.$$
When using the H\'ajek as the base estimator, the aPW \textit{across} method and the \textit{within} method give
$$\hat\theta_{1,\text{H\'ajek-aPW}}:=\frac{\sum_{i=1}^N\bar w_{1i}Y_i}{\sum_{j=1}^N\bar w_{1j}}~~\text{and}~~\hat\theta_{1,\text{H\'ajek-within}}:=\frac{1}{M}\sum_{k=1}^M\hat\theta_{1,\text{H\'ajek-}k}.$$
These two estimators can also be re-expressed as linear outcome combinations,
$$\hat\theta_{1,\text{H\'ajek-aPW}}=\sum_{i=1}^N\frac{\left(\frac{1}{M}\sum_{k=1}^Mw_{1ik}\right)}{\sum_{j=1}^N\left(\frac{1}{M}\sum_{l=1}^Mw_{1jl}\right)}Y_i~~\text{and}~~\hat\theta_{1,\text{H\'ajek-within}}=\sum_{i=1}^N\left(\frac{1}{M}\sum_{k=1}^M\frac{w_{1ik}}{\sum_{j=1}^Nw_{1jk}}\right)Y_i.$$
In both of these linear combinations, the coefficients are formed out of the collection of the $NM$ inverse probability weights $w_{1ik}$. The difference is that, to form the coefficients, in the \textit{within} estimator the weights $w_{1ik}$ are first normed (i.e., scaled to sum to 1) in each imputed dataset and then are averaged across imputed datasets, whereas in the aPW estimator the weights $w_{1ik}$ are first averaged across imputed datasets and then normed. As $N\to\infty$, the sum of the weights in each imputed dataset (and thus the sum of their averages across imputed datasets) approaches $N$, so in both linear combinations the coefficient for unit $i$'s outcome approaches $\frac{\sum_{k=1}^Mw_{1ik}}{NM}$. Thus asymptotically, the H\'ajek-aPW estimator is equivalent to the H\'ajek-within estimator (and also to the H-T estimators), so it is consistent. 

The takeaway is that the aPW method is consistent because it is equivalent to the \textit{within} method if using the Horvitz-Thompson version of inverse probability weighting, and it is asymptotically equivalent to the \textit{within} method if using the H\'ajek (i.e., weighted average) version of inverse probability weighting.

\section{Imputing a function of the missing covariate}

\subsection{Imputing the propensity score (imPS)}\label{sec:imputePS}

The \textit{within} and \textit{across} approaches above rely on imputing $X$. An alternative approach is to treat the propensity score as a variable with missingness -- it can be estimated for units with observed $X$, but is missing for units with unobserved $X$ -- and impute the propensity score.

This approach was first explored by \cite{ling2021HowApplyMultiple}. There are two issues with the specific method in this paper that we need to address before discussing the approach. First, note that the approach requires estimating the propensity score for units with observed $X$ before attempting propensity score imputation for the other units. 
In \cite{ling2021HowApplyMultiple} this initial step is based on a model fit to complete cases, which is problematic because this model estimates $\P(A=1\mid Z,X,R=1)$ which is different from $e(Z,X)=\P(A=1\mid Z,X)$ -- except for the special case where the missingness depends on the observed covariate $Z$ only, $R\independent(X,A,Y)\mid Z$.
This means the method does not actually impute the right propensity score.%
\footnote{Neither \cite{ling2021HowApplyMultiple} nor we make this assumption. It should be noted however that when the missingness in $X$ only depends on $Z$, it is fine to use propensity scores from a complete-case model as the response variable in the imputation model.}
To correct this initial step, we propose weighting complete cases by inverse probability of response, $\frac{1}{\P(R=1\mid Z,A,Y)}$, before fitting the propensity score model. (This weighting is done only to estimate the propensity score model, after which the weights are discarded.) An alternative is to fill in the data with imputed $X^*$ before fitting the propensity score model, but then we might as well use the $X^*$-based propensity scores and would not need to impute propensity scores.
Second, the method in \cite{ling2021HowApplyMultiple} imputes both $X$ and the propensity score in a standard multiple imputation procedure that does not specifically respect the fact that the propensity score is a deterministic function of $(Z,X,A)$, i.e., conditional on these variables there is no randomness in the propensity score. To avoid such model misspecification, we recommend either imputing the propensity score but not imputing $X$ or imputing $X$ and then estimating propensity scores based on that imputation (the conventional approach), but not imputing $X$ and the propensity score together. 

Now we consider the corrected method where the initial propensity score model is appropriate (using inverse probability of response weighting), so the propensity score $e(Z,X)$ is consistently estimated for units with observed $X$, and the multiple imputation is for the missing propensity score only, ignoring $X$. This is what we refer to as the \textit{imputing the propensity score} (imPS) method from this point.

To simplify presentation, let $U:=e(Z,X)$ denote the propensity score. 
Note that $\tau=\E\{\E[Y\mid U,A=1]-\E[Y\mid U,A=0]\}$ because the propensity score is a balancing score \citep{rosenbaum1983CentralRolePropensity}, so $\tau$ is a function of the joint distribution of $(U,A,Y)$.
This means a \textit{within} method using imputation of $U$ instead of $X$ also results in consistent $\tau$ estimation -- under the same condition of correct models so that the imputation recovers this joint distribution.
More broadly, one could impute any one-to-one function of $U$ and then back-transform.

There is a special case with univariate (but not multivariate) $X$ where a choice of function of $U$ has an interesting connection to the original \textit{within} method. If the propensity score model is a generalized linear model, one option is to impute the linear predictor; a rationale is it might be easier to specify the imputation model for the linear predictor than for the propensity score. (To make this concrete, with a logit model, the linear predictor is the logit propensity score.) In this case, if the model terms are $Z,X,A$ and possibly their interactions but no other nonlinear function of $X$, then imputation of the linear predictor is implicitly an imputation of $X$ (because the linear predictor is a linear function of $X$). In this setting, it is better to impute $X$, which helps fill in both $X$ and $U$ for for units with missing data.

\subsection{Mean-imputing the probability weight (imPW)}

When using propensity score weighting as the base method, a different option is to impute the weight $\omega(Z,X,A)$ directly rather than imputing the propensity score; this is consistent as long as the imputation model is mean-correct, i.e., it is correct for $\E[\omega(Z,X,A)\mid Z,A,Y]$. In this case though, one could simply use $\E[\omega(Z,X,A)\mid Z,A,Y]$ as a proxy weight rather than imputing the weight; this is the same proxy targeted by the aPW \textit{across} method. What this means is that a single mean imputation (or prediction) suffices; stochastic imputation is not needed.

\section{Discussion}

We conclude that the \textit{within} method is valid, and can be applied to any type of propensity score analysis or any other effect estimation methods, provided that these base methods are consistent in the no missing data case. A different \textit{within} method that relies on imputing the propensity score rather than the covariate (imPS) is also consistent; it restricts the choice of base estimators to those that do not use covariates once propensity scores are estimated. With the \textit{across} approach, methods that aim to create one set of estimated propensity scores (aPS, aPM) are inconsistent because they fail to achieve a balancing score due to differential measurement error. On the other hand, for propensity score weighting, the aPW \textit{across} method, which uses the average probability weight as proxy weight, is consistent. Related to this, a method that mean-imputes the probability weight given the fully observed variables (imPW) is consistent.

These consistency results are based on the assumption that the imputation model and the analysis model are correctly specified. If the analysis model is incorrect, so that with full data the base estimator would converge to $\tau^*\neq\tau$, and the imputation model is correct, then the methods portrayed as consistent above are no longer consistent for $\tau$ but are consistent for $\tau^*$.

While the paper only covers \textit{within} and \textit{across} methods, we note that a fractional imputation method in \cite{seaman2014InverseProbabilityWeighting} is consistent for the same reason the \textit{within} method is consistent.%
\footnote{This method in \cite{seaman2014InverseProbabilityWeighting} imputes missing values based on a MLE model fit, but this consistency result holds for either MLE-based or Bayesian imputation.}
In this method, the imputed datasets are not analyzed separately but stacked together into one large dataset to be analyzed.

We now turn to consider what the results of this paper mean for practice. This requires combining them with other considerations. One topic not discussed above is the combination of covariate balancing with outcome regression to improve precision or induce robustness. (This may be in different forms, e.g., regression on covariate-balanced data \citep{ho2007MatchingNonparametricPreprocessing}, or using the augmented inverse probability weighting estimator \citep{robins1995AnalysisSemiparametricRegression}.) The \textit{within} method naturally allows for all these method variations, as any method developed for the full-data setting can be applied to imputed datasets. For propensity score weighting analysis using the aPW \textit{across} method, however, there seems to be no straightforward way to combine with an outcome regression. One could construct an \textit{across} method that involves averaging several quantities (the probability weight, the outcome mean function and a product of the two) across imputed datasets, but then that estimator is similar to applying the \textit{within} method to the augmented inverse probability weighting estimator. The imPS method is a \textit{within} method where the imputation recovers the joint distribution of exposure, outcome and the propensity score, so it allows outcome models that adjust for the propensity score but not for the covariates themselves.

Based on consideration of consistency and the general interest in incorporating outcome regression in some form, we would \textit{generally} recommend using the standard \textit{within} method.
If outcome regression is not used, on the other hand, there are a couple of other options: (i) the aPW \textit{across} method can be used, which provides a single set of weights; and (ii) if one believes that the missingness in $X$ depends on $Z$ only, the other option is to estimate the inverse probability weights for units with full data based on a complete-case model and then use the imPW method mean-imputing the probability weight for units with missing covariates.

After judgment of a method's validity (e.g., consistency) come questions about its variance and variance estimation. Variance (i.e., how uncertain estimates from this method are) is a technical topic, as variance depends specifically on how the imputation is done (e.g., Bayesian or MLE-based) and how the imputed data are used \cite[see e.g.,][]{seaman2014InverseProbabilityWeighting}. Variance estimation (i.e., what tools are available to estimate the uncertainty of the estimate), on the other hand, is of practical interest to users. For the \textit{within} method, standard practice is to estimate variance by pooling within- and between-imputation variance using the so-called Rubin's rules \citep{rubin1987multiple}. This is simple to implement \cite[see e.g.,][]{greifer2023EstimatingTreatmentEffects}, provided that one obtains from each imputed dataset an estimate of the effect and of the corresponding standard error.%
\footnote{The example in \cite{greifer2023EstimatingTreatmentEffects} uses within-imputation standard error estimated by Delta method.}
Some authors note that this method relies on correctly specified imputation and analysis models \citep{bartlett2020BootstrapInferenceMultiple}. Variance estimation with mis-specified models or without an estimate of within-imputation variance, often using the bootstrap, is under active research \cite[e.g.,][]{schomaker2018BootstrapInferenceWhen,brand2019CombiningMultipleImputation,bartlett2020BootstrapInferenceMultiple}. There is also a literature on variance estimation when using MLE-based imputation \citep{robins2000InferenceImputationEstimators,seaman2014InverseProbabilityWeighting,vonhippel2021MaximumLikelihoodMultiple}, which involves procedures tailored to the estimand and estimator.

For the aPW and imPW methods, no specific variance estimators have been developed, so this is an open research topic. It might be tempting to estimate variance using survey analysis software, which treats the weights as if they were known (survey) weights. Caution is needed, however, as this may carry both conservative and anti-conservative tendencies: treating estimated inverse propensity score weights as known weights for ATE estimation%
\footnote{This is not true for ATT estimation \citep{reifeis2022VarianceTreatmentEffect}.} is conservative (i.e., overestimates variance) \citep{lunceford2004StratificationWeightingPropensity} but ignoring the fact that some weights are proxy weights is likely anti-conservative (i.e., underestimates variance).

\section*{Acknowledgements}

This work was supported by funding from the National Institute of Mental Health grant R03MH128634 (PI Nguyen). The authors thank three anonymous reviewers and the editor, Dr. Stephen Cole, for helpful comments and suggestions.

\bibliography{zotero}

\appendix

\appendixpage

\section{Revisiting appendix 2b in \cite{leyrat2019PropensityScoreAnalysis}}\label{app:A}

We reproduce here the argument from \cite{leyrat2019PropensityScoreAnalysis} (in our notation) for the claim that $A\independent Y_a\mid Z,X^*$ where $X^*$ is generated from the correct model, i.e., $\P(X^*=x\mid Z,A,Y)=\P(X=x\mid Z,A,Y)$:
\begin{align*}
    \P(A=1\mid Y_a,Z,X^*=x)
    &=\frac{\P(X^*=x\mid A=1,Y_a,Z)\P(A=1\mid Y_a,Z)}{\P(X^*=x\mid Y_a,Z)}
    \\
    &=\frac{\P(X=x\mid A=1,Y_a,Z)\P(A=1\mid Y_a,Z)}{\P(X=x\mid Y_a,Z)}
    \\
    &=\P(A=1\mid Y_a,Z,X=x)
    \\
    &=\P(A=1\mid Z,X=x)
    \\
    &=\P(A=1\mid Z,X^*=x).
\end{align*}
The one mismatch in this proof is in the transition from the denominator in the first line to that in the second line, specifically equating $\P(X^*=x\mid Y_a,Z)$ to $\P(X=x\mid Y_a,Z)$. To see why, apply iterated expectation (or the law of total probability) to both of these densities, obtaining
\begin{align*}
    \P&(X^*=x\mid Y_a,Z)
    \\
    &=\textcolor{blue}{\P(X^*=x\mid Y_a,A=a,Z)}\P(A=a\mid Y_a,Z)+\textcolor{red}{\P(X^*=x\mid Y_a,A=1-a,Z)}\P(A=1-a\mid Y_a,Z),
    \\
    \P&(X=x\mid Y_a,Z)
    \\
    &=\textcolor{blue}{\P(X=x\mid Y_a,A=a,Z)}\P(A=a\mid Y_a,Z)+\textcolor{red}{\P(X=x\mid Y_a,A=1-a,Z)}\P(A=1-a\mid Y_a,Z).
\end{align*}
Note that the two blue terms are equal to each other (due to SUTVA and correct imputation model),
\begin{align*}
    \textcolor{blue}{\P(X^*=x\mid Y_a,A=a,Z)}
    &=\P(X^*=x\mid Y,A=a,Z)
    =\P(X=x\mid Y,A=a,Z)
    =\textcolor{blue}{\P(X=x\mid Y_a,A=a,Z)}.
\end{align*}
However, the two red terms are not equal to each other. We put them in similar forms below to make the difference clear. The red term involving $X^*$ is:
\begin{align}
    \textcolor{red}{\P(X^*=x\,}&\textcolor{red}{\mid Y_a,A=1-a,Z)}\nonumber
    \\
    &=\E[\P(X^*=x\mid Y_a,Y_{1-a},A=1-a,Z)\mid Y_a,A=1-a,Z] && \text{(iterated expectation)}\nonumber
    \\
    &=\E[\P(X^*=x\mid Y_a,Y,A=1-a,Z)\mid Y_a,A=1-a,Z] && \text{(SUTVA)}\nonumber
    \\
    &=\E[\P(X^*=x\mid Y,A=1-a,Z)\mid Y_a,A=1-a,Z] && \text{(imputation conditions on $(Z,A,Y)$)}\nonumber
    \\
    &=\E[\P(X=x\mid Y,A=1-a,Z)\mid Y_a,A=1-a,Z] && \text{(correct imputation model)}\nonumber
    \\
    &=\E[\P(X=x\mid Y_{1-a},A=1-a,Z)\mid Y_a,A=1-a,Z], && \text{(SUTVA)}\label{leyrat-crux-1}
    \intertext{while the red term involving $X$ is:}
    \textcolor{red}{\P(X=x\,}&\textcolor{red}{\mid Y_a,A=1-a,Z)}\nonumber
    \\
    &=\E[\P(X=x\mid Y_a,Y_{1-a},A=1-a,Z)\mid Y_a,A=1-a,Z]. && \text{(iterated expectation)}\label{leyrat-crux-2}
\end{align}
While the expectations in (\ref{leyrat-crux-1}) and (\ref{leyrat-crux-2}) share the same conditioning set, the density of $X$ inside the expectation in (\ref{leyrat-crux-1}) (which conditions on one potential outcome) is different from the density of $X$ inside the expectation in (\ref{leyrat-crux-2}) (which conditions on both potential outcomes).



\section{An alternative explanation of the consistency of the \textit{within} method based on recovering unconfoundedness}\label{app:B}

The explanation of the \textit{within} method's consistency in the main text of the paper (section \ref{sec:within}) starts with equating the ATE to $\tau:=\E\{\E[Y\mid Z,X,A=1]-\E[Y\mid Z,X,A=0]\}$ (a function of the $(Z,X,A,Y)$ joint distribution), and then considers imputation as a way to recover this joint distribution in the imputed datasets, which justifies using the imputed datasets to estimate $\tau$. That is, the problem to be solved by multiple imputation, in that explanation, starts with $\tau$.
The alternative explanation we now present starts with the ATE instead of with $\tau$. 

In the absence of missing data, ATE identification hangs on the unconfoundedness assumption, $A\independent Y_a\mid Z,X$, for $a=0,1$. This means if we have an imputation scheme that recovers the joint distributions of $(Z,X,A,Y_a)$, for $a=0,1$, we recover unconfoundedness, and this would justify using the imputed dataset to estimate the ATE. Recall from section \ref{sec:within}, though, that covariate imputation does not recover these joint distributions, i.e.,
\begin{align}
    \P(Z,X^\dagger=x,A,Y_a)\neq\P(Z,X=x,A,Y_a).\label{norecover}
\end{align}
The way to recover the joint distributions of $(Z,X,A,Y_a)$ by imputation requires a more complex imputation scheme. Fortunately, we will see that one only needs to imagine (but not to implement) this imputation scheme.

In this imputation scheme, we consider the missingness in both covariate $X$ and potential outcomes $Y_a$. This now is a special case of monotone missingness, where the missingness in $X$ depends on observed variables $Z,A,Y$ but not $X$ itself (our original MAR assumption); and the missingness in $Y_a$ depends on $X$ plus the observed variables $Z$ but not $Y_a$ itself (under the unconfoundedness assumption $A\independent Y_a\mid Z,X$). The imputation scheme involves
%
%
first imputing $X^*$ and then imputing $Y_1^*$ and $Y_0^*$ using the following models:
\begin{align*}
    \P(X^*=x\mid Z,A,Y)&\overset{\text{set}}{=}\P(X=x\mid Z,A,Y),
    \\
    \P(Y_a^*=y\mid Z,X^\dagger=x)&\overset{\text{set}}{=}\P(Y_a=y\mid Z,X^\dagger=x)=\P(Y_a=y\mid Z,X=x),~~\text{for}~a=0,1.
\end{align*}
This imputation achieves
\begin{align}
    \P(Z,X^\dagger=x,A,Y)&=\P(Z,X=x,A,Y),
    \\
    \P(Z,X^\dagger=x,A,Y_a^\dagger=y)&=\P(Z,X=x,A,Y_a=y),\label{recover}
\end{align}
i.e., recovery of the joint distribution of $(Z,X,A,Y)$ and the joint distributions of $(Z,X,A,Y_a)$, for $a=0,1$. Note the difference between (\ref{recover}) and (\ref{norecover}).

The recovery of the $(Z,X,A,Y_a)$ joint distributions implies that the imputed and original datasets share the same ATE. Moreover, this distribution recovery combined with unconfoundedness in the original dataset imply unconfoundedness in the imputed dataset, $A\independent Y_a^\dagger\mid Z,X^\dagger$, for $a=0,1$, which justifies using the imputed dataset to estimate the ATE.

Within the imputed dataset, under unconfoundedness the ATE is equal to a parameter $\tau^\dagger$ defined similarly to $\tau$ in the original dataset, so we can estimate $\tau^\dagger$ to estimate the ATE. (It is obvious that $\tau^\dagger=\tau$ due to the equivalence of the distributions.)
Note that in the imputed dataset $Y=AY_1^\dagger+(1-A)Y_0^\dagger$, so $\tau^\dagger:=\E\{\E[Y\mid Z,X^\dagger,A=1]-\E[Y\mid Z,X^\dagger,A=0]\}$, which involves the observed outcome only, not the imputed values of the potential outcomes. 

What this means is that for the sake of estimation, we do not need to actually impute $Y_a$. Because estimation uses variables $(Z,X^\dagger,A,Y)$, imputation of $X$ is sufficient.
Our consideration of this more complex imputation scheme is just a thought exercise, which gives another explanation for why the \textit{within} method works.

\end{document}